\documentclass{article}

\usepackage{graphicx} 
\usepackage{amsmath}
\usepackage{xcolor} 
\usepackage{soul}
\usepackage[margin=1in]{geometry}
\usepackage[square,numbers]{natbib}
\usepackage{authblk}
\usepackage{url}
\usepackage{tabularx}

\usepackage{fancyhdr}
\title{Data complexity signature predicts quantum projected learning benefit for antibiotic resistance}

\author[1,$\ast$]{Kahn Rhrissorrakrai}
\author[1]{Filippo Utro}
\author[2]{Alex Milinovich}
\author[3]{Sandip Vasavada}
\author[2,5]{Daniel Rhoads}
\author[1]{Laxmi Parida}
\author[3,6]{Glenn T. Werneburg}

\affil[1]{IBM Research, IBM T.J. Watson Research Center, Yorktown Heights, NY, USA}
\affil[2]{Quantitative Health Sciences, Cleveland Clinic, Cleveland, OH, USA}
\affil[3]{Department of Urology, Cleveland Clinic, Cleveland, OH, USA}
\affil[4]{Department of Pathology and Laboratory Medicine, Cleveland Clinic, Cleveland, OH, USA}
\affil[5]{Cleveland Clinic Lerner College of Medicine, Case Western Reserve University, Cleveland, OH, USA}
\affil[6]{Present address: Department of Urology, University of Michigan, Ann Arbor, MI, USA}

\date{}
\begin{document}

\maketitle

\section*{Abstract}
This study presents the first large-scale empirical evaluation of quantum machine learning for predicting antibiotic resistance in clinical urine cultures. Antibiotic resistance is amongst the top threats to humanity, and inappropriate antibiotic use is a main driver of resistance. We developed a Quantum Projective Learning (QPL) approach and executed 60 qubit experiments on IBM Eagle and Heron quantum processing units. While QPL did not consistently outperform classical baselines, potentially reflecting current quantum hardware limitations, it did achieve parity or superiority in specific scenarios, notably for the antibiotic nitrofurantoin and selected data splits, revealing that quantum advantage may be data-dependent. Analysis of data complexity measures uncovered a multivariate signature, which comprised Shannon entropy, Fisher Discriminant Ratio, standard deviation of kurtosis, number of low-variance features, and total correlations.  The multivariate model accurately (AUC = 0.88, $p$-value = 0.03) distinguished cases wherein QPL executed on quantum hardware would outperform classical models. This signature suggests that quantum kernels excel in feature spaces with high entropy and structural complexity. These findings point to complexity-driven adaptive model selection as a promising strategy for optimizing hybrid quantum-classical workflows in healthcare. Overall, this investigation marks the first application of quantum machine learning in urology, and in antibiotic resistance prediction. Further, this work highlights conditional quantum utility and introduces a principled approach for leveraging data complexity signatures to guide quantum machine learning deployment in biomedical applications.

\section{Introduction}
In recent years there have been tremendous strides in the advancement of quantum computing technologies~\cite{levine_parallel_2019, burkard_semiconductor_2023, conlon_error_2019, kozhanov_next-generation_2023, bravyi_future_2022, chamberland_building_2022, martinis_optimal_2021} and a corresponding interest in understanding for which applications this technology would be most beneficial~\cite{cerezo_challenges_2022, abbas_challenges_2024, caro_generalization_2022}.  There is a growing body of evidence that for certain problems, such as optimization~\cite{koch_quantum_2025}, simulation~\cite{daley_practical_2022}, and high energy physics~\cite{di_meglio_quantum_2024}, quantum computing may be able to make advancements. However, in the area of quantum machine learning (QML), there remains to understood where quantum
advantage is guaranteed~\cite{cerezo_challenges_2022, cerezo_variational_2021}. This is particularly true in the current era's context of pre-fault tolerant quantum devices (PFTQD), where quantum devices remain prone to error\cite{google_quantum_quantum_2024, bravyi_high-threshold_2024}, and fault tolerant quantum devices (FTQD) of sufficient size to be non-trivial are several years away~\cite{gambetta_expanding_2022, abughanem_ibm_2025}. Despite the lack of discovered theoretical guarantees, there remains the possibility of empirical discovery of certain instances wherein a QML algorithm may yield improved performance, either in terms of accuracy or speed.

Projected quantum kernels (PQKs) have recently attracted attention due to evidence suggesting that, under specific conditions, they can outperform classical kernel methods~\cite{huang2021power}. This approach is particularly appealing because it addresses two critical constraints in quantum machine learning: maximizing predictive accuracy while minimizing the quantum computational overhead associated with parameterized feature maps on PFTQDs. Unlike variational quantum circuits (VQCs) or quantum neural networks (QNNs)~\cite{abbas_power_2021}, PQKs do not require iterative parameter optimization, thereby avoiding issues such as barren plateaus that hinder trainability in variational models. Empirical studies have demonstrated that PQK variants applied to time-series financial data achieved substantial improvements in predictive performance\cite{ciceri_enhanced_2025}. Furthermore, recent benchmarking efforts comparing quantum kernels to classical radial basis function (RBF) kernels across diverse data modalities indicate that quantum kernels can outperform classical counterparts for certain high-dimensional, complex datasets~\cite{jiang_benchmarking_2025}. However, those findings were limited to simulated environments, and practical quantum advantage remains contingent on scaling to quantum devices.

\begin{figure*}[!t]
\centering
\includegraphics[width=0.9\textwidth]{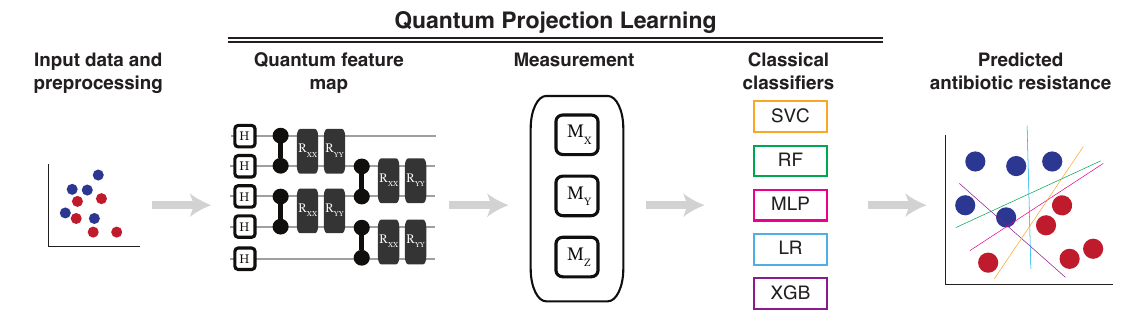}
\caption{Overview of quantum projection learning workflow. Input data are preprocessed classically and may include normalization, dimensionality reduction, and feature selection. After data are processed, they are quantum projected with a quantum feature map.  Measurements are then made in the X, Y, and Z basis. This measured projected data are then analyzed by a suite of classical machine learning classifiers: support vector classifier (SVC), random forest (RF), multilayer perceptron (MLP), logistic regression (LR), and extreme gradient boosting (XGB). These classifiers perform the final antibiotic resistance prediction on the measured data. }
\label{figure:overview}
\end{figure*}

In this study, we investigate the applicability of quantum projection learning (QPL) to a real-world biological dataset. Our QPL takes inspiration from PQK where the quantum projected data measured after embedding by a quantum feature map is analyzed by not only a classical kernel method, but four other classical ML models. A primary objective here is to characterize the conditions under which QPL achieves superior performance relative to classical algorithms. To this end, we conduct experiments on both quantum simulators and quantum devices, enabling us to assess the impact of data characteristics on predictive performance on PFTQDs. 

As a clinically relevant task, we focus on the antibiotic resistance profile of clinically-obtained urine cultures, where accurate and rapid prediction is critical due to the growing prevalence of antimicrobial resistance, the urgency of initiating effective treatment, and the inherent delays associated with culture-based diagnostic tests~\cite{werneburg_external_2025}. 
Antibiotic resistance is associated with more than 4 million deaths globally each year, and is amongst the top threats to humanity today~\cite{walsh_antimicrobial_2023}.  Early prediction of a culture’s resistance and sensitivity to clinically-relevant antibiotics represents a marked opportunity to improve antibiotic selection and reduce selective pressure for resistant organisms, thus with great potential to improve patient outcomes and future resistance profiles at the individual and population levels. 
While traditional machine learning approaches have demonstrated reasonable predictive accuracy using patient clinical data, their performance varies across antibiotics, highlighting the need for improved models. Furthermore, these methods can be challenged by the severe class imbalance present in the data~\cite{johnson_survey_2019}, as resistant cases for specific antibiotics are often underrepresented, complicating model generalization.

These considerations underscore the importance of empirically evaluating quantum machine learning approaches on clinically relevant, real-world datasets. By focusing on antibiotic resistance prediction, this study aims to provide insight into the practical conditions under which quantum projection learning may offer an advantage over classical methods. Our work not only benchmarks QPL against established machine learning algorithms on both simulated and physical quantum hardware, but also seeks to identify the data properties that influence performance in the current pre-fault-tolerant quantum regime. In doing so, we contribute to bridging the gap between theoretical promise and applied utility in quantum machine learning for healthcare.

\section{Methods}

\subsection{Data preparation}\label{methods:data_prep}

\begin{figure}[!t]
\centering
\includegraphics[width=\columnwidth]{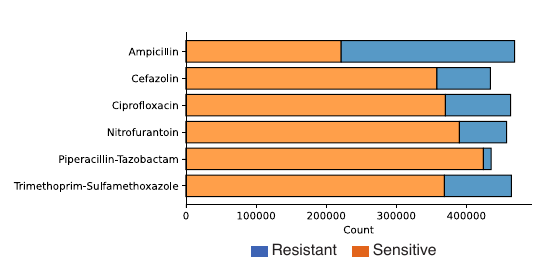}
\caption{Number of organism–antibiotic susceptibility classifications for each antibiotic.}
\label{figure:datasets}
\end{figure}

The clinical data comprised 2,723,116 organism–antibiotic susceptibility classifications derived from six antibiotics (Fig.~\ref{figure:datasets}). Full details are provided in the referenced study~\cite{werneburg_external_2025}.
This data contained 192 categorical and 24 continuous variables.  Categorical variables were binary encoded using the \textit{category-encoders} packages and \textit{BinaryEncoder}.  For each antibiotic, CORDS2 was applied to generate representative coresets using the \textit{cords} library with the \textit{GradMatchStrategy} function. We specified as the model to CORDS2 a simple feedforward neural network with two fully connected layers. The first layer maps the input features to a 512-dimensional embedding. The second layer is the output layer, which produces predictions for two classes. We specified a cross entropy loss and \textit{selection\_type} = 'PerClass'.  For each antibiotic, we generated 10 coresets collections with each collection containing a 300 sample training set and 75 sample testing set.

Dimensionality reduction was performed using NMF, PCA, or UMAP. NMF was performed using the \textit{NMF} function from the \textit{scikit-learn} package with \textit{max\_} = 200 after the data first min-max scaled using \textit{MinMaxScaler} from \textit{scikit-learn}. PCA was performed using the \textit{PCA} function from \textit{scikit-learn}  after the data scaled using \textit{StandardScaler} from \textit{scikit-learn}. UMAP was performed using the \textit{UMAP} function from the \textit{umap} package after the data scaled using \textit{StandardScaler}.

\subsection{Quantum projection learning}\label{methods:QP}
\subsubsection{Quantum projection}
The quantum projection approach begins by encoding classical data into a quantum circuit, a process referred to as quantum embedding. This embedding maps input features into a high-dimensional Hilbert space through parameterized quantum gates, enabling the representation of complex correlations that are difficult to capture classically. After state preparation, the quantum state is projected back into classical space via measurement. Specifically, measurements are performed in all single-qubit Pauli bases to construct the complete set of single-qubit reduced density matrices (1-RDMs). These 1-RDMs provide expectation values that serve as transformed features, forming a new dataset on which classical learning methods are applied.

For quantum circuit construction, we adopted embedding strategies proposed by Huang et al.~\cite{huang2021power}, with the primary embedding implemented as the ZZ Feature Map~\cite{havlicek_supervised_2019}, widely used in quantum machine learning for its ability to encode pairwise feature interactions through entangling gates. This feature map applies parameterized rotations followed by controlled-Z entanglement, generating a non-linear mapping of input data into the quantum state space. The secondary quantum embedding that we implement is a Hamiltonian evolution ansatz. In particular, we use the 1D-Heisenberg model Hamiltonian with 4 or 6 Trotter steps as our quantum embedding.

For the quantum projection experiments, projective measurements in the X, Y and Z basis, which determine the corresponding expectation values, were made. The 1-RDM for each qubit can then be determined by combining these expectation values. Therefore, we carry out three different experiments (one for each measurement basis) for each data point. In each experiment, we used 10,000 shots, i.e. repeated executions of the quantum circuit, to calculate a statically significant estimate of the expectation value. 

The experiments were performed on two IBM Eagle R3 quantum processing units (QPUs) (\textit{ibm\_cleveland}-pre-August 2025 and \textit{ibm\_brussels}) and two IBM Heron R2 QPUs (\textit{ibm\_aachen} and \textit{ibm\_cleveland}-post-August 2025). All are superconducting qubit devices with fixed frequency transmon qubits as data qubits connected via tunable couplers. To improve the quality of the obtained results we have utilized a combination of error suppression and mitigation techniques. Pauli twirling was used to tailor the noise in gate and measurement operations to Pauli noise. 

\subsubsection{Machine learning methods}\label{methods:ml_methods}
For the following ML methods applied to the quantum projected data, the hyperparameters were optimized via \textit{RandomizedSearchCV} using five-fold cross validation over 40 iterations.

\paragraph{Support Vector Classifier}
Support Vector Classifier is a supervised learning algorithm that constructs a maximum-margin hyperplane to separate classes in feature space, optionally using kernel functions to enable non-linear decision boundaries. SVC is effective for high-dimensional data and robust to overfitting when regularization is properly tuned. We implement this using the \textit{SVC} function from \textit{scikit-learn} and optimized hyperparameters over the following ranges: \textit{C} (0.1, 1, 10, 100), \textit{gamma} (0.001, 0.01, 0.1, 1), and \textit{kernel} (linear, rbf, poly, sigmoid).

\paragraph{Random Forest}
Random Forest (RF) is an ensemble method that aggregates predictions from multiple decision trees trained on bootstrap samples with randomized feature selection at each split. This de-correlation improves generalization and reduces variance. RF is well-suited for tabular data with complex interactions. We optimized hyperparameters over the following ranges: \textit{n\_estimators} (100–900 in increments of 100), \textit{max\_depth} (5–19), \textit{min\_samples\_split} (2–9), \textit{min\_samples\_leaf} (1–4), and \textit{bootstrap} (True or False).

\paragraph{Multilayer perceptron}
MLP is a feedforward neural network composed of multiple layers of neurons with non-linear activation functions, trained using backpropagation and gradient-based optimization. It captures non-linear relationships and interactions between features. We implemented this using the \textit{MLPClassifer} function from \textit{scikit-learn} and optimized hyperparameters over the following ranges: \textit{hidden\_layer\_sizes} ((128,64,32,10), (64,32,10), (128,64,32)), \textit{activation} (identity, logistic, tanh, relu), \textit{solver} (lbfgs, sgd, adam), and \textit{alpha} (0.00005, 0.0005).

\paragraph{Logistic regression}
Logistic Regression models the probability of class membership using a logistic function applied to a linear combination of input features. Regularization terms (L1, L2) are often included to prevent overfitting and handle multicollinearity. We implemented this using the \textit{LogisticRegression} function from \textit{scikit-learn} and optimized hyperparameters over the following ranges: \textit{C} (0.001, 0.01, 0.1, 1, 10, 100), \textit{penalty} (l1, l2), and \textit{solver} (liblinear, saga).

\paragraph{Extreme gradient boosting}
Extreme Gradient Boosting (XGBoost or XGB) is a gradient boosting algorithm that builds decision trees sequentially, where each tree minimizes residual errors from previous iterations using gradient-based optimization. It incorporates regularization and advanced optimizations for scalability and performance. We optimized hyperparameters over the following ranges: \textit{n\_estimators} (100, 200, 300), \textit{learning\_rate} (0.01, 0.1, 0.2), \textit{max\_depth} (3, 5, 7), \textit{subsample} (0.7, 0.8, 1.0), \textit{colsample\_bytree} (0.7, 0.8, 1.0), and \textit{min\_child\_weight} (1, 3, 5).

\subsection{Geometric separation }
The original formulation of the PQK method introduced a theoretical framework for assessing the potential for quantum advantage in machine learning tasks. This framework was primarily designed for evaluating classical kernel methods in conjunction with quantum kernels. In our study, only one of the five classical baselines employs a kernel-based approach; however, for completeness and comparability, we implemented a similar evaluation strategy. Specifically, we define a metric $g_{cq}$, which quantifies the geometric separation between kernels constructed from classical embeddings and those derived from quantum-projected embeddings.

\begin{equation}
    g_{cq}=g(K^c||K^q)=\sqrt{||\sqrt{K^q}\sqrt{K^c}(K^c+\lambda I)^{-2}\sqrt{K^q}\sqrt{K^c}||_\infty}
\end{equation}

In the above equation, $\lambda$ is the regularization parameter in the primal formalism of the SVM, and $K$ are the classical and quantum-projected kernel matrices. If this quantity is on the order of $\sqrt{N}$, where $N$ is the number of the training samples in the dataset (in our study N=300), then we can move on to the second test to check for potential quantum prediction advantage. If $g_{cq}$  is significantly smaller than $\sqrt{N}$, then we can expect the classical model to perform as well as the quantum-projected model.

For the second test, we define additional metrics referred to as model complexities $s_c$ and $s_q$. A potential quantum prediction advantage can be expected in case $s_c$ is on the order of $\sqrt{N}$, while $s_q$ is smaller. Model complexity is defined as follows.

\begin{align}
    s_{K,\lambda}(N)=\sqrt{\frac{\lambda^2\sum_{i=1}^{N}\sum_{j=1}^{N}{\left(K+\lambda I\right)_{ij}^{-2}y_iy_j}}{N}} + \nonumber \\
    \sqrt{\frac{\sum_{i=1}^{N}\sum_{j=1}^{N}{\left(\left(K+\lambda I\right)^{-1}K\left(K+\lambda I\right)^{-1}\right)_{ij}y_iy_j}}{N}}
\end{align}

In the above equation, all symbols are as defined previously, and $y$ are the true labels of the dataset. 

We performed an initial evaluation of our dataset focused on Ampicillin with these metrics using the first two motif positions with binary encoding and eight repetitions of the ZZ Feature Map . The results reveal a geometric separation between the classical and the quantum-projected radial basis function kernels of 14.833, which is similar to $\sqrt{N} = 17.321$. In the second check, we observed the classical model complexity to be 0.6681 vs 1.3249 for the quantum-projected kernel. Considering a different antibiotic representative of the class-imbalanced antibiotics, the separation was only 0.0079, though the complexity of the classical and quantum models were 1.000 and 0.447.  For both antibiotics, only one of the two conditions for potential advantages were met, though they did not agree on the met conditions.

\subsection{Data complexity measures}
We selected a wide range of data complexity measures to characterize input and embedded datasets. Please see Table S1 for a summary.

\subsection{Statistical testing of association between complexity and QPL performance}
To identify a multivariate signature of data complexity measures predictive of QPL performance, we employed a logistic regression model with an elastic net penalty, which combines $L_1$ and $L_2$ regularization to balance feature sparsity and stability. The model was implemented using scikit-learn’s \textit{LogisticRegressionCV} with the \textit{saga} solver, which supports elastic net optimization for high-dimensional datasets. Hyperparameter tuning included \textit{l1\_ratio} $= \{0.25, 0.5, 0.75\}$ to control the relative contribution of $L_1$ and $L_2$ penalties. Model evaluation was performed using stratified 5-fold cross-validation. Recursive Feature Elimination (RFE) was applied to rank and select the most informative complexity measures, yielding a reduced feature set that maximized predictive performance. The final model was assessed using area under the ROC curve (AUC) and statistical significance testing to confirm discriminative capability.

\subsection{Classical machine learning baselines}\label{methods:basline}

\paragraph{Random Forest}
Random Forest is an ensemble learning algorithm that extends bagging by introducing randomness in feature selection to reduce tree correlation and improve generalization~\cite{breiman_random_2001}. Each tree is trained on a bootstrap sample, and splits consider random feature subsets, with predictions aggregated by majority voting. RF is effective for high-dimensional tabular data and complex feature interactions, outperforming many deep learning models in such settings~\cite{grinsztajn_why_2022}, though performance may degrade with small datasets or highly correlated features. We implemented RF using \textit{scikit-learn} (v1.6.1) and optimized hyperparameters via \textit{RandomizedSearchCV} over the following ranges: \textit{n\_estimators} (100–900 in steps of 100), \textit{max\_depth} (5–19), \textit{min\_samples\_split} (2–9), \textit{min\_samples\_leaf} (1–4), and \textit{bootstrap} (True or False).

\paragraph*{Extreme gradient boosting}
Extreme Gradient Boosting is a gradient boosting algorithm that builds decision trees sequentially, with each tree minimizing residual errors from prior iterations using gradient-based optimization~\cite{chen_xgboost_2016}. It incorporates $L_1$ and $L_2$ regularization to reduce overfitting and employs optimizations such as tree pruning, sparsity-aware split finding, and cache-efficient memory access for scalability. Compared to Random Forests, XGBoost often achieves superior generalization but is more sensitive to hyperparameter tuning. We implemented XGBoost using \textit{XGBClassifier} from the \textit{xgboost} library (v3.0.0) and optimized hyperparameters via \textit{GridSearchCV} in scikit-learn (v1.6.1),over the following hyperparameter values: \textit{n\_estimators} (100, 200), \textit{max\_depth} (3, 5, 7), \textit{learning\_rate} (0.01, 0.1, 0.2), subsample (0.7, 0.8, 1.0), \textit{colsample\_bytree} (0.7, 0.8, 1.0), and \textit{min\_child\_weight} (1, 3, 5).

\section{Results}\label{results}
This study represents the first application of quantum machine learning in urology, and in antibiotic resistance prediction. From individuals with available culture sensitivity data at our healthcare enterprise, we have clinical data with 216 features as input to our hybrid quantum-classical classifier, as well as their antibiotic sensitivity and resistance classification as response label. We focused on six clinically relevant antibiotics: Ampicillin, Cefazolin, Ciprofloxacin, Nitrofurantoin, Trimethoprim-Sulfamethoxazole, and Piperacillin-Tazobactam. For each of these antibiotics, we have 100s of 1000s of samples with all but Ampicillin being highly class imbalanced with respect to antibiotic resistance (Fig.~\ref{figure:overview}A).

\subsection{Hybrid quantum-classical workflow}\label{workflow}
We implemented a hybrid quantum–classical framework termed Quantum Projective Learning (QPL), which applies a quantum projection of classical clinical input data into a high-dimensional Hilbert space, followed by training multiple classical machine learning (ML) models on the projected representation (Fig.~\ref{figure:overview}B). This approach extends the concept of Projected Quantum Kernels (PQKs), originally introduced by Huang et al.~\cite{huang2021power}, which leverages quantum feature maps to enhance expressivity for classical datasets. We perform extensive parameter testing considering entanglement type, feature map repetition, number of features, and dimensionality reduction techniques. While PQK-based methods typically pair quantum embeddings with kernel-based learners such as Support Vector Classifiers (SVCs), we broadened this paradigm by incorporating additional classical learners, including Random Forests (RF), Multi-Layer Perceptrons (MLP), Logistic Regression (LR), and Extreme Gradient Boosting (XGB), to evaluate whether non-kernel models can exploit the quantum embedded feature space. All classical models underwent systematic hyperparameter optimization using grid and Bayesian search strategies. This design aims to assess whether the increased expressivity afforded by quantum projections can improve predictive performance beyond kernel-based approaches~\cite{ciceri_enhanced_2025}.

\begin{figure*}[!t]
\centering
\includegraphics[width=0.7\textwidth]{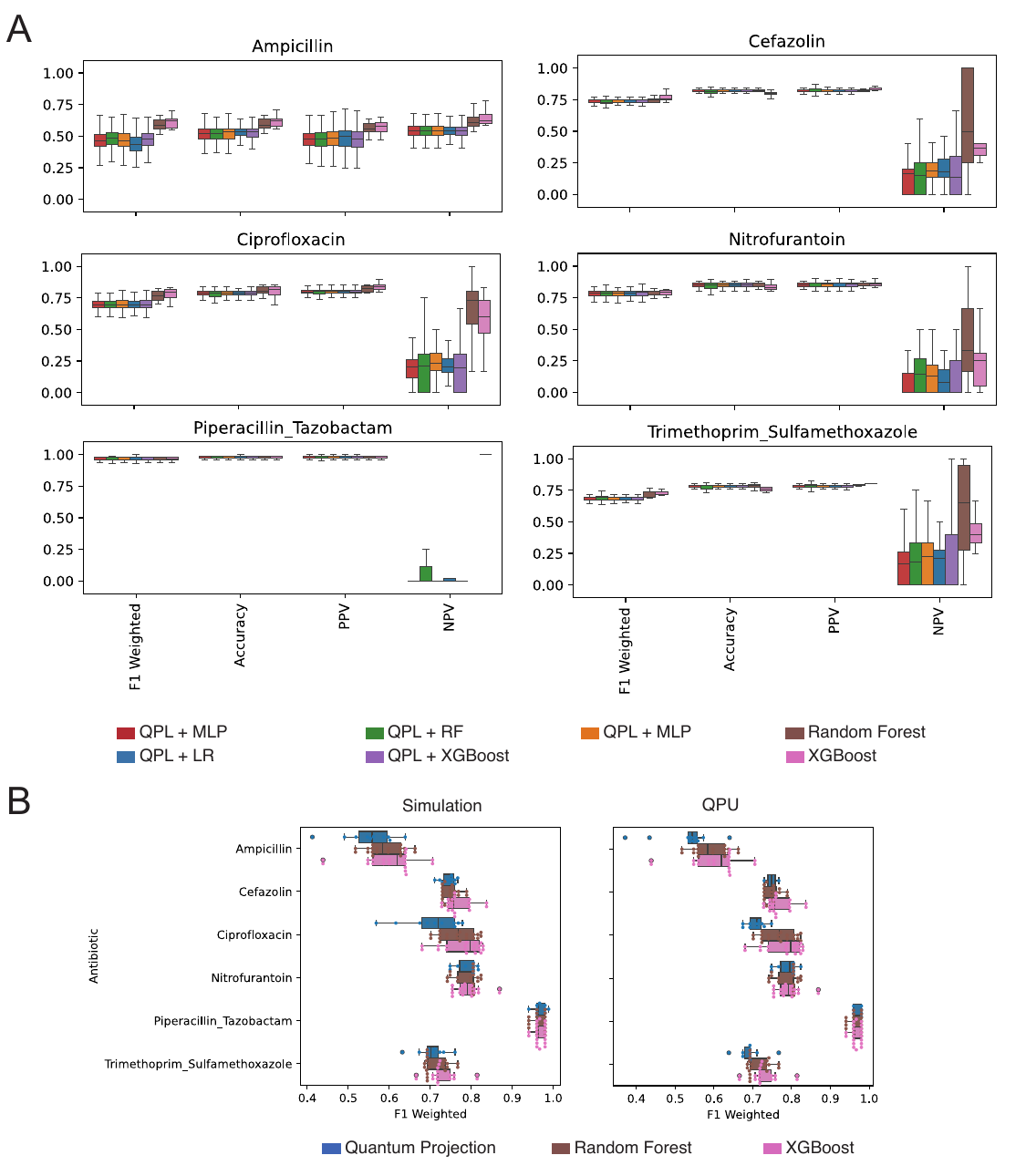}
\caption{Antibiotic resistance prediction performance per antibiotic. A) Predictive performance across four metrics for each of the five QPL types and two classical baselines. B) Performance comparison between the best performing model configuration for QPL, RF, and XGB executed on simulator or QPU for each data split. The best performing model configuration is the model with the maximum median weighted F1 over 10 data splits.}
\label{figure:best_overall}
\end{figure*}

\subsection{Capturing characteristics of data subsets}\label{data_features}
Although the available clinical dataset is extensive, the current size, connectivity and error rates of current quantum hardware necessitate dimensionality reduction to maintain feasible circuit widths and depths for PFTQDs, where noise significantly impacts computational fidelity. Additionally, given the relatively low clock speeds of QPUs, efficient resource utilization requires constraining the number of samples processed. This constraint arises because each circuit corresponds to a single data sample and must be executed multiple times (i.e., “shots”) to estimate output probability distributions accurately. To address these limitations, we employed CORDS2~\cite{killamsetty2021retrieve} to generate representative subsets (“coresets”) of the original dataset, creating 10 stratified train/test splits per antibiotic, with training sets of size 300 and test sets of size 75.

We assessed the coresets using a suite of data complexity measures to quantify structural and distributional differences across antibiotics (Supplementary Table 1, Fig. S1). Metrics exhibiting low variability (variance $\leq 0.001$ across all samples) were excluded. Overall, most complexity measures demonstrated minimal variation across antibiotics, indicating broadly similar feature space characteristics. However, Shannon entropy exhibited a notable divergence: ampicillin displayed substantially higher entropy, suggesting greater class distribution uncertainty and increased randomness, whereas trimethoprim-sulfamethoxazole exhibited markedly lower entropy, indicative of more deterministic class structure. Additionally, trimethoprim-sulfamethoxazole consistently presented a higher proportion of low-variance features, implying reduced feature informativeness and potential redundancy within its representation. While minor variability was observed among individual train/test splits within each antibiotic, these differences were limited, supporting that CORDS2 effectively generated representative and internally consistent coresets across all antibiotics.

\subsection{Predictive performance per antibiotic}\label{per_antibiotic} 
We systematically evaluated QPL across an extensive parameter space using a noiseless state vector simulator, and on selected configurations on a quantum processing unit (QPU) (Fig.~\ref{figure:best_overall}). The QPU experiments involved circuits encoding over 60 input features, necessitating quantum circuits with $\geq 60$ qubits. Our objective in testing first in simulation was to quantify the impact of key design choices, including dimensionality-reduction strategies, entanglement topology (e.g., linear vs. pairwise), feature map repetition , and the number of encoded input dimensions. In total, we conducted 43,202 simulation-based experiments (Fig. S2) to explore these factors systematically. Five QPL variants were tested, each combining quantum projection with a distinct classical classifier: SVC, RF, MLP, LR, and XGB. RF and XGB were also evaluated as standalone classical baselines to benchmark performance against QPL approaches.

\begin{figure*}[!t]
\centering
\includegraphics[width=0.9\textwidth]{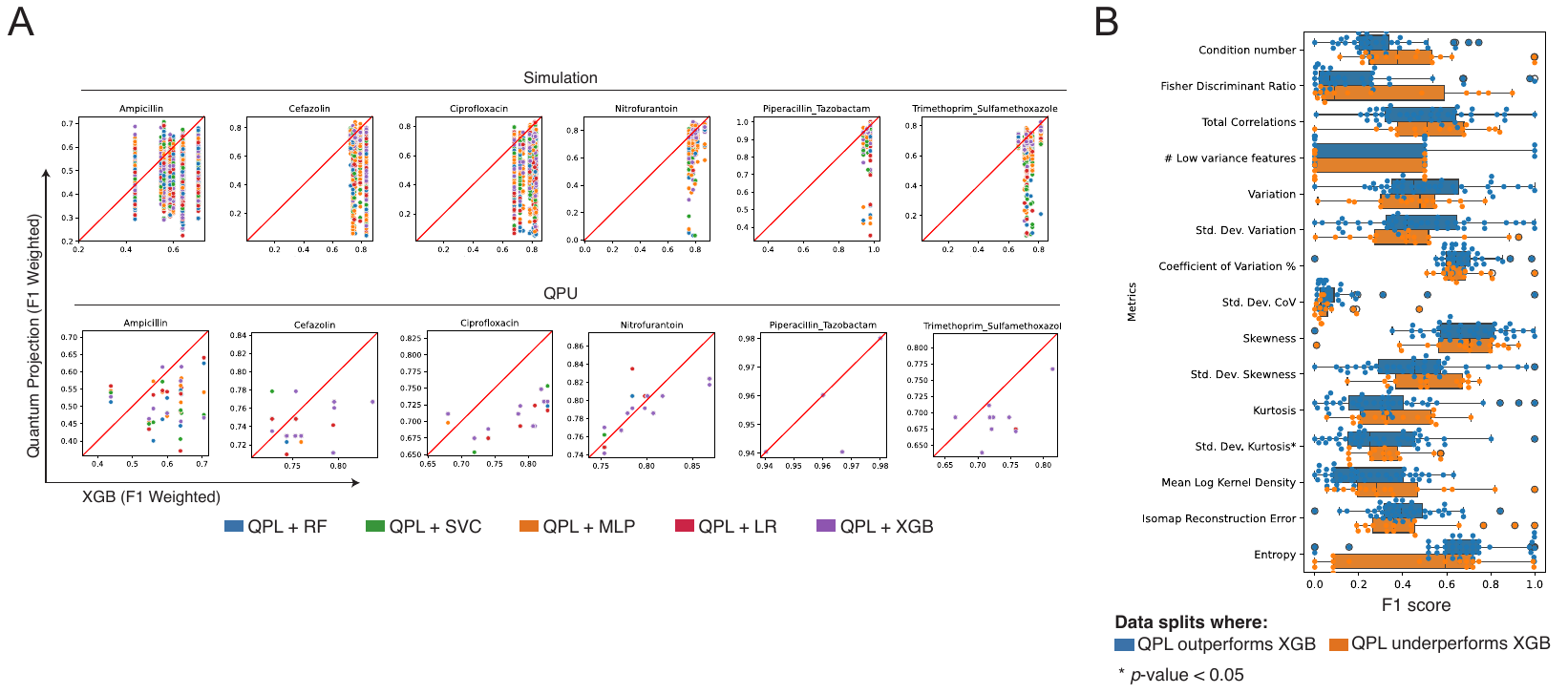}
\caption{Predictive performance of QPL versus XGBoost in simulation and hardware.  A) Datapoints represent individual data splits across all QPL types for each antibiotics. Weighted F1 of the QPL and XGB are on the y- and x-axis, respectively. Experiments are delineated by QPLs executed on simulator (top) and on QPU (bottom). B) Comparison of the data complexity metric distribution between data splits where QPL outperforms (blue) or under performs (orange) XGB in simulation.
}\label{figure:best_splits_XGB}
\end{figure*}

Given that dimensionality reduction is often required to align feature space size with the constraints of PFTQDs, we systematically evaluated the impact of three widely used techniques: Principal Component Analysis (PCA), Uniform Manifold Approximation and Projection (UMAP), and Non-Negative Matrix Factorization (NMF). These methods were applied prior to quantum feature mapping to assess their influence on downstream model performance. While using the original, non-embedded feature set generally yielded the highest performance, this configuration was only computationally feasible for classical baselines (RF and XGB) (Fig. S2). Among the dimensionality reduction methods, NMF exhibited the greatest variability across experiments and consistently underperformed relative to PCA and UMAP. PCA and UMAP demonstrated comparable performance profiles, with both slightly outperforming NMF across most antibiotics and parameter settings. These findings suggest that linear (PCA) and manifold-based (UMAP) embeddings preserve discriminative structure more effectively than factorization-based approaches in this context, while also enabling circuit sizes compatible with current quantum hardware limitations.

We restricted entanglement strategies to pairwise and linear entanglement rather than full entanglement. While full entanglement can theoretically enhance expressivity by enabling global correlations across all qubits, it significantly increases gate count and error susceptibility, making it impractical for PFTQDs. In contrast, pairwise and linear entanglement provide a controlled mechanism for distributing correlations while maintaining more manageable circuit depth and error rates. Across all experiments, we observed no statistically significant performance difference between pairwise and linear entanglement configurations (Fig. S2), suggesting that for the tested problem sizes and feature maps, additional connectivity beyond linear chains does not confer measurable advantage under current hardware constraints.

Analogous to entanglement strategies, the choice of feature map repetition represents a trade-off between circuit expressivity and practical constraints such as circuit depth, gate count, and error accumulation~\cite{daspal_effect_2023}. Increasing repetitions theoretically enhances the representational capacity of the quantum feature map by introducing additional layers of parameterized rotations and entanglement, thereby enabling the encoding of higher-order correlations among input features. However, this comes at the cost of increased circuit depth. In our experiments, varying the number of feature map repetitions across the tested range yielded negligible differences in predictive performance (Fig. S2). This observation suggests that, for the problem sizes and feature distributions considered, additional repetitions do not confer a measurable benefit and that shallow circuits with minimal repetition may achieve comparable accuracy while reducing resource overhead and error accumulation.

We further examined the effect of varying the number of input features derived from embeddings generated by NMF, UMAP, and PCA. Feature subsets of size 2-16 and 60 were evaluated to quantify the relationship between input dimensionality and model performance. Quantum hardware experiments for 60 feature experiments were performed on two IBM Eagle R3 QPUs (\textit{ibm\_cleveland}-pre-August 2025, \textit{ibm\_brussels}) and two IBM Heron R2 QPUs (\textit{ibm\_aachen} and \textit{ibm\_cleveland}-post-August 2025). These high-dimensional, hardware tests were primarily conducted using UMAP embeddings. Across most antibiotics, there was no statistically significant difference in predictive accuracy between 2 and 16 features, suggesting that low-dimensional embeddings capture similarly discriminative structure for these cases (Fig. S2). However, for cefazolin and trimethoprim-sulfamethoxazole, we observed a modest performance improvement when using 60 features, indicating that higher-dimensional representations may preserve additional relevant variance for certain antibiotics. 

\begin{figure*}[!t]
\centering
\includegraphics[width=0.8\textwidth]{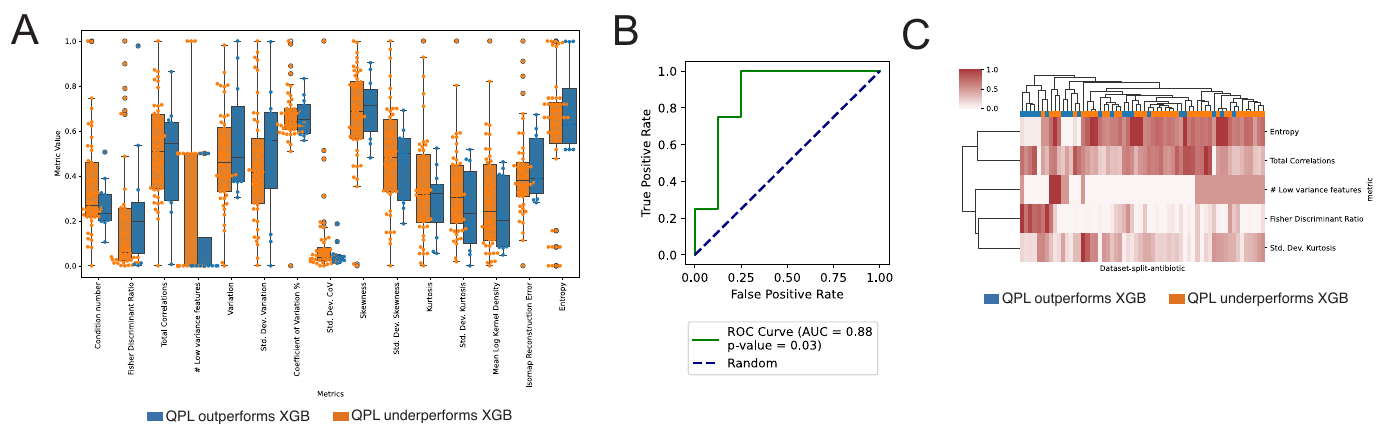}
\caption{Relationship of data complexity measures and QPL performance on QPU. A) Boxplot of distribution of metrics values for data splits where QPL outperformed (blue) and underperformed (orange) XGB on IBM Heron R2 QPUs. B) AUC for correctly predicting QPL performance using a logistic regression with ElasticNet penalty in a five fold cross validation. C) Heatmap of metric values found significant by logistic regression. Rows are significant metrics and rows antibiotic data splits. Column color bar indicates whether the QPL outperformed (blue) or underperformed (orange)  XGB}\label{figure:best_Qfeatures_hardware}
\end{figure*}

Across a total of 43,202 simulation-based experiments and 472 experiments based on hardware executions, we observed that RF and XGB outperformed QPL for the majority of antibiotics across multiple evaluation metrics (Fig.~\ref{figure:best_overall}A). For the classification task with the most balanced class distribution (ampicillin), QPL achieved a median $F_1$ score and accuracy of approximately 0.50, whereas RF and XGB attained median values near 0.70, indicating a substantial performance gap under these conditions. For the remaining five antibiotics, all methods demonstrated improved $F_1$ score, accuracy, and positive predictive value (PPV), albeit with a corresponding reduction in negative predictive value (NPV), reflecting the impact of class imbalance on predictive reliability. Notably, the performance differential between QPL and classical models was less pronounced for these antibiotics, and in the case of Nitrofurantoin, QPL achieved parity with both RF and XGB, suggesting that hybrid quantum-classical approaches may offer competitive performance for certain data distributions.

Given the large parameter space explored for QPL configurations, we next identified the best-performing parameter settings for each learning method, QPL, RF, and XGB, based on the maximum median $F_1$ score aggregated across all train/test splits for each antibiotic (Fig.~\ref{figure:best_overall}B). This approach allowed us to isolate optimal configurations under both simulated and hardware-executed conditions. Consistent with earlier observations, RF and XGB continued to outperform QPL for ampicillin, ciprofloxacin, and trimethoprim-sulfamethoxazole, achieving higher median $F_1$ scores and accuracy across splits. For the remaining antibiotics, performance differences between QPL and classical baselines were less pronounced, with several cases exhibiting near-parity. Furthermore, we found there not be a significant degradation in performance for experiments performed on QPU and the relative performance patterns were indeed maintained from simulation to hardware execution. Moreover, there has been empirical evidence showing that a Heisenberg evolution circuit rather than a ZZ feature map may be beneficial to projected quantum kernel approaches~\cite{huang2021power,ciceri_enhanced_2025}. We tested several configurations of this circuit for antibiotics piperacillin-tazobactam and nitrofurantoin at 4 and 6 trotter steps. We found no meaningful difference in performance.

\subsection{Data characteristics associated positive quantum projective performance}\label{quantum_win}

Although QPL did not demonstrate overall improved predictive performance compared to classical baselines, there were specific antibiotic data splits where QPL models outperformed both RF and XGB (Fig.S3, \ref{figure:best_splits_XGB}A). To investigate potential drivers of these localized improvements, we analyzed whether particular data characteristics differentiated overperforming versus underperforming splits for QPL, under both simulation and hardware execution conditions. Among the 15 evaluated data complexity measures, two exhibited statistically significant divergence between these groups: standard deviation of kurtosis ($p$-value = 0.027) and coefficient of variation percentage ($p$-value = 0.031), associated with XGB and RF comparisons respectively (Fig.~\ref{figure:best_splits_XGB}B, S4). These measures suggest that variability in feature dispersion and distributional shape may influence QPL’s relative advantage. Additional metrics, including Shannon entropy, Isomap reconstruction error, and kurtosis, showed observable deviations but did not reach significance thresholds. Collectively, these findings indicate that QPL performance gains may be linked to specific structural properties of the feature space rather than global dataset characteristics.

To specifically assess QPL performance under hardware-executed conditions, where circuits encoded a substantially larger feature space (60 dimensions) compared to simulation-based experiments (maximum of 16 dimensions), we examined whether data complexity measures could differentiate overperforming from underperforming splits. No single measure exhibited statistically significant divergence between these groups (Fig.~\ref{figure:best_Qfeatures_hardware}A). However, when applying a logistic regression model with an elastic net penalty to the full set of complexity metrics, we achieved robust separation between over- and underperforming splits, yielding an AUC of 0.88 and a $p$-value of 0.03 (Fig.~\ref{figure:best_Qfeatures_hardware}B). A similar regression model applied to all experiments, on both quantum hardware and simulator, did not yield a significant separation. This suggests the presence of a multivariate signature predictive of QPL performance only under high-dimensional conditions afforded by hardware-execution. The five most influential features identified by the model were Shannon entropy, Fisher Discriminant Ratio, Standard Deviation of Kurtosis, Number of Low-Variance Features, and Total Correlations (Fig.~\ref{figure:best_Qfeatures_hardware}C). These measures collectively capture aspects of class separability, distributional shape, feature redundancy, and overall complexity, indicating that structural properties of the feature space may inform when QPL achieves competitive or superior performance on quantum hardware.

\section{Discussion}\label{sec12}
This study represents the first large-scale empirical evaluation of quantum machine learning for predicting antibiotic resistance in in clinical urine cultures using a hybrid quantum–classical framework, Quantum Projective Learning. Our findings provide critical insights into the practical feasibility, performance characteristics, and limitations of quantum-enhanced models in clinical predictive tasks under current hardware constraints.

Across more than 43,000 simulation-based experiments and multiple executions on IBM Eagle and Heron quantum processors, QPL did not consistently outperform classical baselines such as Random Forest and Extreme Gradient Boosting. This observation is consistent with recent benchmarking studies, which report that quantum kernels often achieve performance comparable to classical kernels when hyperparameters are optimized, particularly in PFTQDs where noise, limited qubit connectivity, and shallow circuit requirements constrain expressivity~\cite{jiang_benchmarking_2025, alvarez-estevez_benchmarking_2025}. These results underscore the challenge of realizing theoretical quantum advantage in practical machine learning tasks under current hardware limitations.

Despite the absence of global performance gains, QPL achieved parity or superiority in specific scenarios, notably for nitrofurantoin and selected data splits of other antibiotics. Analysis of data complexity measures revealed that subsets characterized by higher Shannon entropy and greater variability in kurtosis were more likely to benefit from QPL. This suggests that quantum advantage may be data-dependent, emerging in feature spaces with structural properties that are difficult for classical models to capture. These findings align with theoretical predictions that quantum kernels excel when data exhibit non-linear correlations or high-dimensional manifolds~\cite{havlicek_supervised_2019, huang2021power, schuld_quantum_2019}. Furthermore, logistic regression analysis identified a multivariate signature comprising entropy, Fisher Discriminant Ratio, standard deviation of kurtosis, number of low-variance features, and total correlations that discriminates between over- and underperforming QPL splits on hardware with high accuracy (AUC = 0.88). This indicates that complexity-driven adaptive model selection could be a viable strategy for optimizing hybrid quantum-classical workflows.

From an implementation perspective, dimensionality reduction techniques (PCA, UMAP) and circuit design choices (entanglement topology, feature map repetitions) exhibited minimal influence on predictive performance, suggesting that simpler configurations may suffice in QPL. Importantly, the absence of significant degradation when transitioning from simulation to hardware demonstrates that noise-resilient quantum embeddings are achievable, provided circuit depth and gate count remain controlled. However, experiments requiring 60-dimensional embeddings highlight the computational and architectural challenges of scaling quantum circuits, reinforcing the need for efficient encoding strategies and error mitigation.

Several limitations warrant consideration. Hardware constraints restricted exploration of larger feature spaces and more expressive circuits, limiting our ability to fully characterize scaling behavior. While alternative feature maps, including Heisenberg evolution circuits, were tested, no substantial gains were observed. Future work should investigate optimized trotterization schemes and advanced error mitigation techniques. Additionally, our analysis focused on binary classification for antibiotic resistance, yet extending QPL to multi-class classification, survival analysis, or longitudinal prediction tasks could reveal additional insights into quantum model applicability in healthcare.

\section{Conclusion}\label{sec13}
While we did not detect an overall improvement of QPL over classical models in the current dataset, the competitive performance in specific scenarios, and the identification of signatures that predict when QPL is superior to classical models, present a nuanced path toward quantum utility in healthcare. These findings contribute to the growing body of evidence that quantum machine learning, even with pre-fault tolerant hardware, can offer practical benefits when applied judiciously to structured, high-dimensional biomedical data. As quantum hardware and hybrid algorithms continue to mature, the ability to prospectively identify such favorable regimes may play a crucial role in guiding early, high-impact applications of quantum technology in biomedical data analysis and clinical decision support.

\renewcommand{\thefigure}{S\arabic{figure}}
\renewcommand{\thetable}{S\arabic{table}}
\renewcommand{\theequation}{S\arabic{equation}}

\clearpage

\section{Supplementary materials}

\begin{table}[ht]
\centering
\renewcommand{\arraystretch}{1.2} 
\begin{tabular}{|>{\centering\arraybackslash}m{3cm}|>{\centering\arraybackslash}m{6.5cm}|>{\centering\arraybackslash}m{6.5cm}|}
\hline
\textbf{Measure} & \textbf{Description} & \textbf{Modeling Implications} \\
\hline
\# Features & Total number of variables or attributes in the dataset, representing its dimensionality. & High feature count may require dimensionality reduction. \\
\hline
\# Samples & Number of observations or data points in the dataset. & Low sample size relative to features can lead to poor generalization. \\
\hline
Feature Samples ratio & Ratio of features to samples, indicating whether the dataset is high-dimensional relative to its size. & High ratio suggests risk of overfitting. \\
\hline
Intrinsic Dimension & Estimate of the true underlying dimensionality of the data manifold. & Guides choice of dimensionality reduction techniques (e.g., PCA, manifold learning). \\
\hline
Condition number & Measure of numerical stability; high values indicate multicollinearity. & Multicollinearity can destabilize linear models. \\
\hline
Fisher Discriminant Ratio & Quantifies class separability by comparing between-class variance to within-class variance. & Low ratio implies classes are hard to separate. \\
\hline
Total Correlations & Captures redundancy among features by measuring overall statistical dependence. & High redundancy suggests feature selection or decorrelation methods. \\
\hline
Mutual information & Indicates how much information one variable provides about another. & Helps identify informative features for supervised learning. \\
\hline
\# Non-zero entries & Count of non-zero values in the dataset, often used to assess sparsity. & Sparse data may benefit from specialized models. \\
\hline
\# Low variance features & Number of features with minimal variability. & Low-variance features often add noise. \\
\hline
Variation & Average spread of feature values, typically measured as variance. & High variation may require normalization or scaling. \\
\hline
Std. Dev. Variation & Variability of variance across features. & Uneven variance suggests need for standardization. \\
\hline
Coefficient of Variation \% & Ratio of standard deviation to mean (percentage). & High CoV indicates instability. \\
\hline
Std. Dev. CoV & Dispersion of coefficient of variation across features. & Large dispersion may require robust scaling. \\
\hline
Skewness & Measures asymmetry in feature distributions. & Strong skewness may require transformations . \\
\hline
Std. Dev. Skewness & Variability of skewness across features. & Indicates heterogeneity. \\
\hline
Kurtosis & Describes the tails of feature distributions. & High kurtosis suggests outliers. \\
\hline
Std. Dev. Kurtosis & Variation in kurtosis across features. & Large variation may require feature-specific preprocessing. \\
\hline
Mean Log Kernel Density & Average log-density of data points under a kernel density estimate. & Low density may indicate sparse regions. \\
\hline
Isomap Reconstruction Error & Measures how well Isomap preserves distances. & High error suggests complex manifold. \\
\hline
Fractal dimension & Estimates dataset complexity via self-similarity. & Higher complexity may require flexible models. \\
\hline
Entropy & Represents uncertainty or disorder in feature distributions. & High entropy implies diverse data. \\
\hline
Std. Dev. Entropy & Variability of entropy across features. & Uneven uncertainty suggests selective feature weighting or pruning. \\
\hline
\end{tabular}
\caption{Descriptions and Modeling Implications of Data Complexity Measures}
\label{tab:data_complexity_implications}
\end{table}

\clearpage

\begin{figure*}[!t]
\centering
\includegraphics[width=0.9\textwidth]{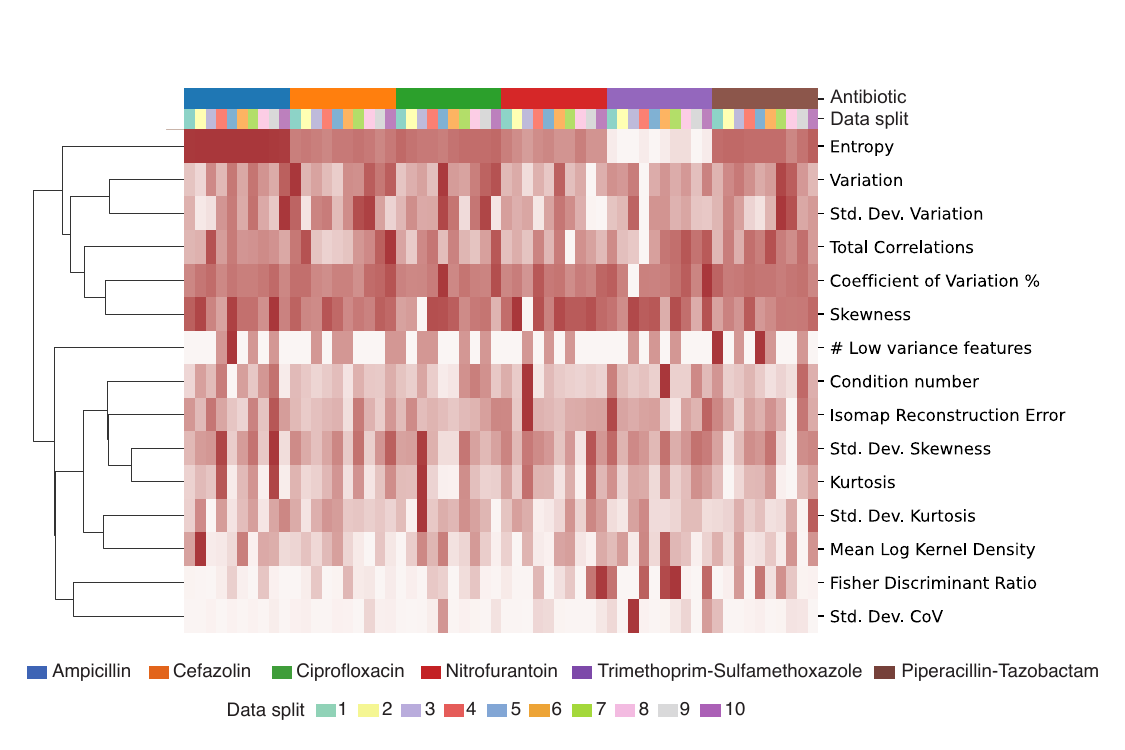}
\caption{Antibiotic resistance data over 10 data splits according to 15 data complexity measures. Metric values are row-wise min-max scaled for purposes of visualization. }
\label{figure:data_metrics}
\end{figure*}

\begin{figure*}[!t]
\centering
\includegraphics[width=0.9\textwidth]{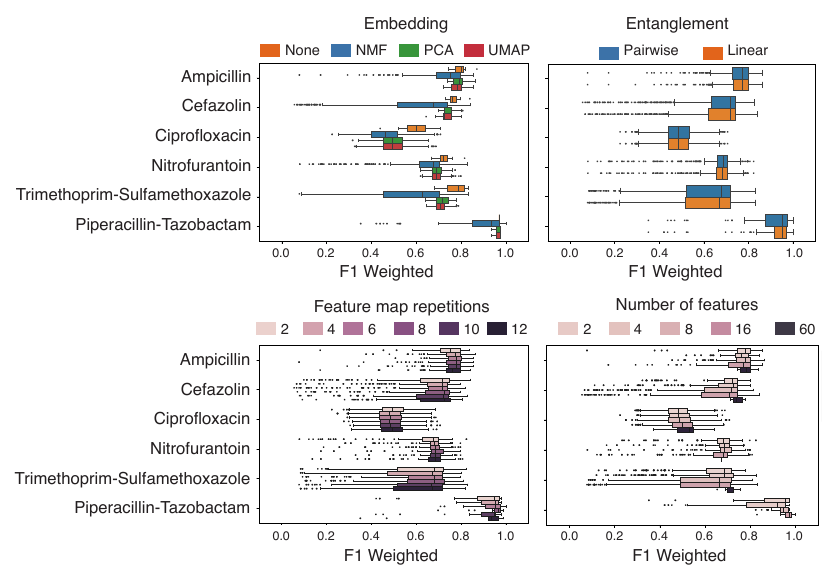}
\caption{Prediction performance across antibiotics in simulation.}
\label{figure:sim_params}
\end{figure*}

\begin{figure*}[!t]
\centering
\includegraphics[width=0.9\textwidth]{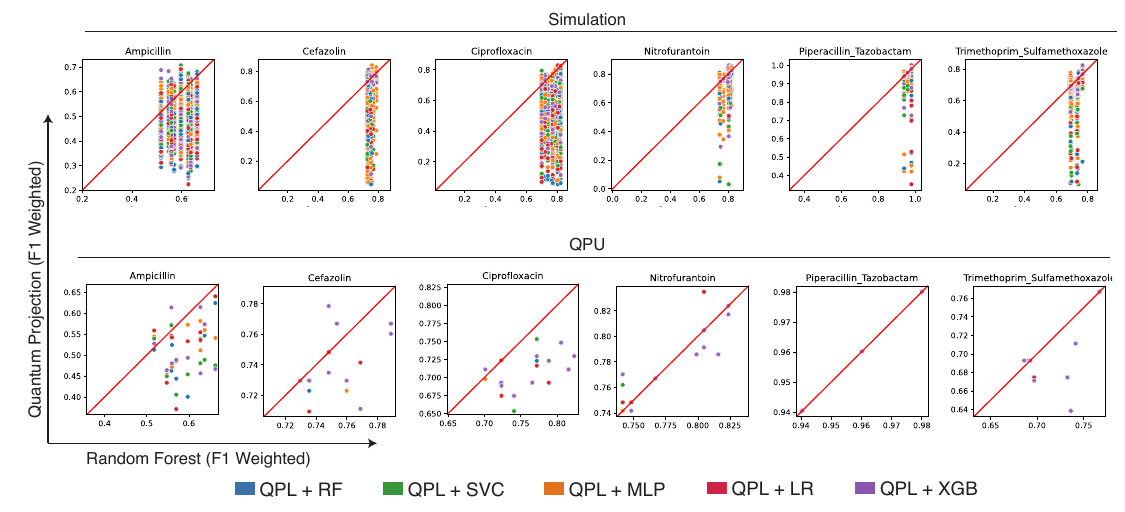}
\caption{Predictive performance of QPL versus Random Forest.  Datapoints represent individual data splits across all QPL types for each antibiotics. Weighted F1 of the QPL and RF are on the y- and x-axis, respectively. Experiments are delineated by QPLs executed on simulator (top) and on QPU (bottom)}\label{figure:best_splits_RF}
\end{figure*}

\begin{figure*}[!t]
\centering
\includegraphics[width=0.9\textwidth]{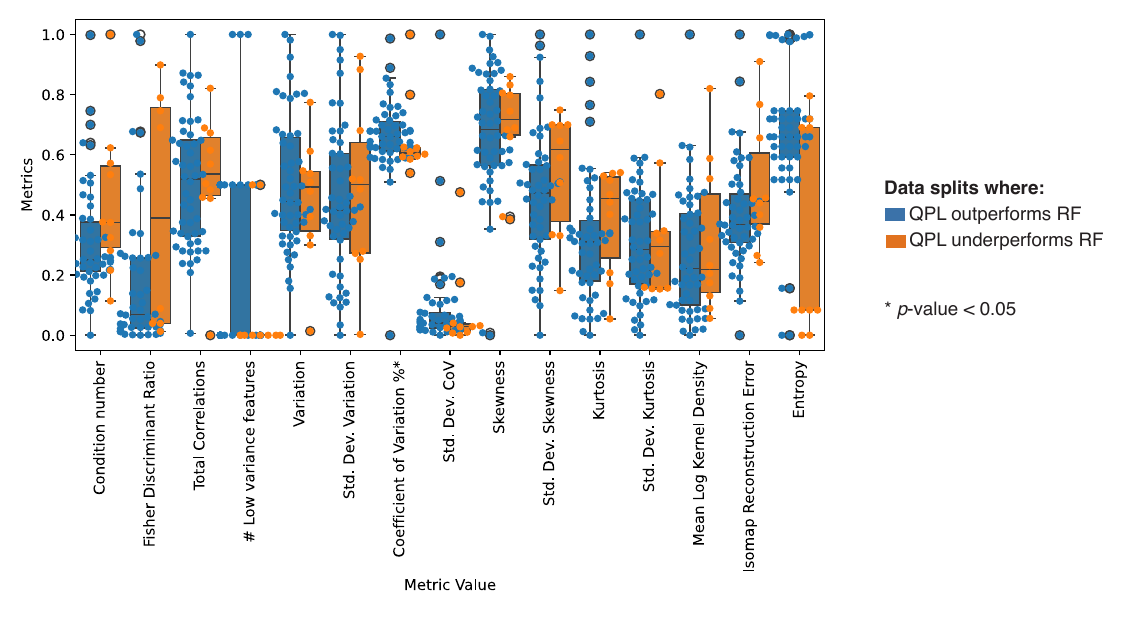}
\caption{Comparison of the data complexity metric distribution between data splits where QPL outperforms (blue) or under performs (orange) RF.
}\label{figure:Boxplot_diff_RF}
\end{figure*}

\clearpage

\bibliographystyle{unsrt}
\bibliography{reference}

\end{document}